# Structural correlations in the enhancement of ferroelectric property of Sr doped BaTiO$_3$

Satish Yadav[1], Mohit Chandra[1], R. Rawat[1], Vasant Sathe[1], A. K. Sinha[2] and Kiran Singh[*,1,3]

[1]UGC-DAE Consortium for Scientific Research, University Campus, Khandwa Road, Indore 452001, India.
[2]HXAL, Synchrotrons Utilization Section, RRCAT, Indore 452013, India
[3]Department of Physics, Dr. B. R. Ambedkar National Institute of Technology, Jalandhar, 144011, India

## Abstract

The effect of Sr doping in BaTiO$_3$ (BTO) with nominal compositions Ba$_{0.80}$Sr$_{0.20}$TiO$_3$ (BSTO) have been explored in its structural, lattice vibration, dielectric, ferroelectric and electrocaloric properties. The temperature dependent dielectric results elucidate the enhancement in dielectric constant and exhibit three frequency independent transitions around 335, 250 and 185 K which are related to different structural transitions. All these transitions occur at lower temperature as compared with pristine BTO, however; remnant electric polarization (*P*) of BSTO is much higher than in BTO. The value of *P* is ~ 5μC/cm$^2$ at room temperature and the maximum *P* ~8μC/cm$^2$ is observed at tetragonal to orthorhombic and orthorhombic to rhombohedral transitions. The electro-caloric effect shows the maximum adiabatic change in temperature ΔT ~0.24 K at cubic to tetragonal transition. The temperature dependent synchrotron X-ray diffraction and Raman results shows correlations between *P*, crystal structure and lattice vibrations. Our results demonstrate the enhancement in ferroelectric properties of BTO with Sr doping. The origin of the enhancement in ferroelectric property is also discussed which is related to the appearance of superlattice peak around room temperature due to TiO$_6$ octahedral distortion. These enhanced properties would be useful to design lead free high quality ferroelectric and piezoelectric materials.

[*]**Corresponding Author:** Dr. Kiran Singh

Department of Physics, Dr. B. R. Ambedkar National Institute of Technology, Jalandhar-144011, India

**Email:** singhkp@nitj.ac.in;
  kpatyal@gmail.com

## I. INTRODUCTION

Ferroelectric materials are one of the most important class of functional materials which show switchable spontaneous electric polarization by external electric field. Because of the switchable nature of polarization, these materials have enormous applications in technological devices like in memory, switching, storage devices etc. [1]. Among ferroelectric materials, $BaTiO_3$ (BTO) is the most studied ferroelectric material of perovskite family. It has three structural phase transitions; cubic to tetragonal around 403 K, tetragonal to orthorhombic around 278 K and orthorhombic to rhombohedral around 198 K [2]. The ferroelectricity in pure BTO is driven by the displacement of $Ti^{4+}$ to off centered position and the direction of polarization is along [001], [110] and [111] crystallographic directions in tetragonal, orthorhombic and rhombohedral phases, respectively [3].

The discovery of BTO is the result of the search of high dielectric constant material during Second World War, almost 80 years ago, but still an interesting material with lots of potential applications [4-6]. Recently, BTO and its derivatives have attracted enormous attention because of the increasing functionality of such materials and the demand of lead free ferroelectric materials. It is realized that BTO and its derivatives are the best alternative to the lead based ferroelectric materials [7-8] and are environmental friendly. Although ferroelectricity in the BTO is well known, still its structural instability, nature of all its transitions and effect of different dopants at A and B sites are not completely understood. For example, recently, Ranjan's group have found that large piezo-coefficient at room temperature is the result of coexistence of monoclinic and orthorhombic phases along with tetragonal phase [9]. Besides three structural phase transitions, it is also reported that this system has domain relaxation across each phase transitions due to the formation and rearrangement of the mesoscopic sub-domain and domain boundaries [10].

It is well known that BTO does not exhibit good electric polarization around room temperature. In order to enhance the ferroelectric and dielectric properties of BTO, different types of dopants at A and B sites have been studied. Although the ferroelectric property in BTO is mainly governed by $Ti^{4+}$ off centering, the A site doping can also modulate its ferroelectricity and hence the effect of Sr and Ca doping at A site is explored on its dielectric and ferroelectric properties [11-12]. The origin of the enhancement of ferroelectric properties with Sr doping is not clear. Recently, such materials are also studied for large electro-caloric coefficient [13] e.g. electrocaloric coefficient is enhanced by Sn doping in BTO [14].

In the present work, we report detail temperature dependent structural, Raman scattering, dielectric, ferroelectric and electrocaloric properties of $Ba_{0.80}Sr_{0.20}TiO_3$ (BSTO) across all the transitions. We have also addressed the underlying mechanism responsible for the enhanced ferroelectric properties. Our room temperature XRD results infer that the crystal structure of BSTO is not pure tetragonal but is accompanied with a superlattice peak. In addition to different transitions, our dielectric results corroborate the relaxation behavior within 260 to 330 K (orthorhombic and tetragonal phases). The electric polarization is enhanced in BSTO sample and remnant polarization is around 5 $\mu C/cm^2$ at 280 K. We have also discussed the origin of enhancement of dielectric and ferroelectric properties of BSTO. The temperature dependent electrocaloric property shows the maximum adiabatic change in temperature $\Delta T$ ~0.24 K for electric field change of 500 kV/m at cubic to tetragonal transition. The temperature dependent synchrotron XRD (SXRD) and Raman scattering measurements established the structural and phonon correlations to its dielectric and ferroelectric properties.

## II. EXPERIMENTAL DETAILS

The polycrystalline samples of BSTO were prepared by conventional solid state reaction method as reported previously [15]. The room temperature X-ray diffraction (XRD) measurement is performed on powder sample using D2 Phaser Desktop Bruker X-ray diffractometer with Cu- Kα radiation. The temperature dependent (90 to 370 K) dielectric measurements are performed at different frequencies during cooling and warming (1K/min) using home-made insert coupled with Keysight LCR meter model E4980A and Lakeshore temperature controller [16]. The temperature dependent P-E loops measurements has been carried out using Radiant technology precision multiferroic P-E loop tracer with home-made attachment for temperature variation [17]. The pyroelectric measurement is performed using Keithley electrometer 6517B. In pyroelectric measurement, first the sample is cooled to 90 K in the presence of pooling electric field 500 (kV/m). At 90 K, the electric field is removed and both the surfaces of the sample are short circuited for half an hour to remove the stray charge or any other extrinsic effects. Then current *vs* time measurement is performed at 90 K for one hour to stabilize the current. Finally, current *vs* temperature measurement is performed till 360 K during warming (3K/min). The electric polarization is calculated from the integration of pyrocurrent. To know the temperature variation of electrocaloric effect, it is essential to have the heat capacity as a function of temperature. For this we have developed the home made set up and data acquisition software made on LabVIEW platform based on semi adiabatic heat pulse method as reported earlier [18] . The temperature-dependent SXRD measurements were performed at angle dispersive X-ray diffraction beamline (BL-12) at Indus-2, RRCAT Indore ($\lambda$ = 0.7525 Å). The temperature dependent Raman measurements were carried out using Horiba JY make HR-800 single spectrometer equipped with a 473 nm excitation source, an 1800 g/mm grating and a CCD detector with overall spectral resolution of ~ 1 cm$^{-1}$.

## III. RESULTS AND DISCUSSIONS

### A. Room temperature structure

The room temperature (RT) XRD (Lab. source) pattern of BSTO shows that at room temperature almost all the peaks belong to the tetragonal symmetry. The Reitveld refinement of XRD data using Full Prof software [19] with tetragonal space group *P4mm* is shown in Fig. 1(a). Besides the main tetragonal peaks, one can also observe a peak around 2$\theta$ = 38°, which could not be fitted with tetragonal space group *P4mm*. The zoomed view of this peak is shown in the inset of Fig. 1 (a). We have also checked for the presence of other impurity phases but this is not the case and this additional peak is called as superlattice peak. The appearance of such superlattice peaks are the result of out of phase octahedral (TiO$_6$) distortion in neighboring pseudo unit cell which results the increment of the unit cell size [20-21]. We have used different symmetry to fit this superlattice peak and the best fit is obtained for $R\bar{3}c$ space group. Fig. 1(b) shows the Rietveld refinement of the XRD data with mixed phases *P4mm* and $R\bar{3}c$ space group and the inset shows the zoomed view near the superlattice peak. The similar superlattice peak is also reported in other such system [22]. Our XRD results illustrates that the crystal structure of BSTO at RT is not pure tetragonal *P4mm* but it is the mixture of tetragonal *P4mm* and rhombohedral $R\bar{3}c$. The evaluation of the superlattice peak with temperature will be discussed in the coming section of temperature dependent SXRD results.

## B. Temperature dependent dielectric behaviour

The temperature dependent real part of complex dielectric and tan$\delta$ at different frequencies during warming cycle is shown in Fig. 2 (a) and 2 (b), respectively. Fig. 2 clearly shows three frequency independent transitions around 185, 250 and 335 K. Similar to pristine BTO, all these transitions exhibit hysteretic behavior during cooling and warming cycles which infer the *first order* nature of these transitions. The hysteretic behavior in real part of dielectric constant during cooling and warming cycle at tetragonal to orthorhombic transition (250 K) is shown in the inset of Fig. 2 (a). Like pristine BTO, these peaks correspond to structural phase transitions i.e. rhombohedral to orthorhombic, orthorhombic to tetragonal and tetragonal to cubic from low temperature to high temperature. However, the transitions are shifted towards lower temperature as compared to BTO, which is consistent with earlier work [23]. Interestingly, the value of dielectric constant at cubic to tetragonal transition is around 7000 at 5 kHz which is much higher than the pure BTO [24] and tan$\delta$ value is also less. This shows that the dielectric constant is enhanced for 20% Sr doped sample.

The dielectric constant is a complex quantity and the imaginary part is very important to understand the detail behavior of sample like dipole dynamics, relaxation etc. In most of the earlier reports [12, 23] and [25], the tan$\delta$ as a function of temperature at different frequencies is missing. Besides three frequency independent anomalies in real part and tan$\delta$, we have also observed the frequency dispersion along with broad frequency dependent maximum in tan$\delta$ around 280 to 330 K. The Fig. 2 (b) illustrates that the dispersion is maximum at lower frequencies and visible up to 10 kHz in orthorhombic and tetragonal region. The dispersion in tan$\delta$ infers the relaxation behavior concomitant with long range ordering. This feature is not very clear in the real part of complex dielectric. This relaxation behavior is analyzed using frequency dependent maximum in tan$\delta$. This frequency dependent maxima is fitted with Arrhenius equation and presented in the inset of Fig. 2(b). The activation energy is found to be (0.85 ± 0.05) eV. Such type of relaxation behavior is also observed in pure BTO but at very low frequencies (in mHz frequency range) and it is attributed to the mesoscale subdomain/domain wall relaxation [26, 10]. Our results show that in BSTO, such relaxation is visible even at higher frequencies (20 kHz). Microscopic origin of this relaxation behavior is till now not clear, but probably it would be due to the different orientations of polarization in different phases. As our XRD results infer that BSTO does not have pure *P4mm* phase and most probably it is due to the crystallographic phase coexistence of *P4mm* and $R\bar{3}c$ phases as reported by Kalyani et al. [9]. We will also discuss this feature with the evolution of superlattice peak in the temperature dependent SXRD results. From these results it is clear that the higher value of dielectric around 280 K and relaxation behavior is due to the appearance of this superlattice peak around this temperature regime i.e. the presence of *P4mm* and $R\bar{3}c$ phases. Such enhancement in dielectric constant is also reported for different such materials [22].

### C. Temperature dependent P-E loops, pyroelectric and electrocaloric properties

The isothermal P-E loops at some selected temperatures in different regime for BSTO are shown in Fig. 3(a). Here, one can clearly observe that the shape of P-E loop, saturation polarization, remnant polarization (P) is changing with temperature. At higher temperature i.e. at 380 K, loop is completely disappeared which corresponds to the paraelectric regime of BSTO. The P-E loops of BSTO are much better than the P-E loops of polycrystalline BTO. For direct comparison, the P-E loops of BTO and BSTO at RT are shown in the inset of Fig. 3(a). The value of P in pure BTO and BSTO at RT are found to be ~1.6 µC/cm$^2$ and ~5.6 µC/cm$^2$. This shows that the P in BSTO is around 3.7 times higher than BTO. These features infer the better ferroelectric properties of BSTO sample as compared to BTO.

To further understand the temperature evolution of polarization, the P at different temperatures is extracted from isothermal P-E loops at different temperatures ($E_{max}$=1000V/mm) and is shown in the Fig. 3(b). Inset of Fig. 3(b) shows the P-E loop of BSTO measured at different maximum voltages at RT. Here, it is important to show that at higher field, the saturation polarization increases slightly, whereas, P does not change very much. From Fig. 3(b), one can clearly see all transitions in the temperature dependent P. Interestingly, the P is the highest at the rhombohedral to orthorhombic and orthorhombic to tetragonal phase transitions. The maximum value of P is ~ 8µC/cm$^2$. This temperature behavior of P is similar to the temperature dependence of P along a - axis in single crystal [3].

To further cross check the temperature dependence of P, we have performed pyroelectric measurement as mentioned in the experimental section. The temperature dependent P from pyroelectric measurement is shown in Fig. 4(a). From this figure one can also notice that polarization is maximum across rhombohedral to orthorhombic and orthorhombic to tetragonal phase transitions. Hence, we have confirmed that the careful pyroelectric measurements can provide the similar spontaneous polarization behavior as observed from P-E loops.

These days ferroelectric materials have also attracted attention due to their electrocaloric properties. The electrocaloric effect is the adiabatic change in temperature or isothermal change in entropy with change in applied electric field. For any material, it is characterized by two parameters; adiabatic temperature change ($\Delta T$) and isothermal entropy change ($\Delta S$). One can calculate ($\Delta T$) and ($\Delta S$) using Maxwell equation, given below:

$$\Delta T = -(1/\rho)\int_{E_1}^{E_2} (T/C(T)_p)(\frac{\partial P}{\partial T})_E dE , \quad \Delta S = -(1/\rho)\int_{E_1}^{E_2} (\frac{\partial P}{\partial T})_E dE$$

where $\rho$ is the density and it is found to be 5.1949(3) g/c.c., $T$ is temperature, $C(T)_p$ is specific heat at constant pressure, $P$ is the polarization and $E_1$ and $E_2$ are initial and final electric field.

In many reports, the electrocaloric coefficient (ECC) $(\partial P/\partial T)_E$ is extracted from the P-E loops at different temperatures [14] and [27- 28]. Here, instead of calculating ECE from P-E loops, we have measured it directly using pyrocurrent measurements and assumed that field change is 500 kV/m ($E_2 - E_1$ = poling field). Fig. 4(b) shows the temperature dependent specific heat of the BSTO from 120 to 350 K, where one can observe three clear peaks associated with all the transitions. The inset of Fig. 4(b) shows the clear peak in specific heat associated with rhombohedral to orthorhombic transition. Figs. 4(c) and 4(d) show the temperature dependent $\Delta T$ and $\Delta S$. These figures show that $\Delta T$ and $\Delta S$ are maximum around tetragonal to cubic transition. The maximum value of $\Delta T$ and $\Delta S$ is found to be around 0.24 K and 0.40 (J/kg-K) for 500kV/m electric field change. This is also consistence with other reports [28]. In Fig. 4(a) one observe that polarization is decreasing with temperature in rhombohedral and orthogonal phases and its result comes out as inverse electrocaoric effect in these regions as shown in Fig. 4(b) and Fig. 4(c). The origin of this behaviour could be related to the polycrystalline nature of the sample.

### D. Temperature dependent synchrotron diffraction

The temperature dependent SXRD patterns at some selective temperatures from 200 to 350 K are shown in Fig. 5(a). As we have mentioned in the above section that at RT the crystal structure is not pure tetragonal but has some superlattice peaks. The appearance of superlattice peaks is temperature dependent which can be seen clearly in the temperature range 280 to 335 K. The inset of Fig. 5(a) shows the superlattice peak before (111) peak of tetragonal phase at 320 K. The observed peak positions are indexed according to the perovskite structure [12]. Fig. 5(b) shows the evolution of the {200} peaks with temperature. At 200 K, (200) peak is asymmetric and on the right side it has another peak. This is the result of three peaks in orthorhombic phase corresponding to (200), (020) and (002) peaks. At temperature 250 K, one can clearly observe that peak asymmetric of {200} peaks has been diminished and have only two peaks, which corresponds to the structural transition around 250 K i.e. orthorhombic to tetragonal phase transition. With increasing temperature i.e. at 280 K one can clearly observe that one of the (200) peak is diminishing and completely disappears at 300 K and peak remains asymmetric till 335 K. It is important to mention here that the presence of superlattice peak persist till 335 K. The evolution of this superlattice peak at some selected temperatures is shown in Fig. 5(c). The inset in 280 K pattern illustrates the emergence of this superlattice peak at this temperature and indicated by *. The intensity of this superlattice peak increases with increasing temperature till 335 K. At 350 K in the cubic phase this superlattice peak disappeared completely. The effect of this superlattice peak has the direct correlation with the sharp increase in dielectric constant and electric polarization around the same temperature regime. The change in polarization around 280 K where slope is changing drastically is shown in the shaded region in Fig. 4a. In addition to it, the maxima in dielectric relaxation is also observed around this temperature region. (See Fig. 2b highlighted region).

### E. Temperature dependent Raman scattering

Raman scattering is very useful technique to understand the structural transformation and local symmetry and even very small distortion can be detected at molecular level. Moreover, the phase co-existence in $Ba_{1-x}Ca_xZr_{0.95}Ti_{0.05}O_3$ is also evident by using temperature dependent Raman studies [29]. The RT Raman spectra of polycrystalline BTO and BSTO is presented in Fig. 6(a). At RT, Raman modes of BSTO and pure BTO are identical except for the position of wave numbers and spectra at lower wave number i.e. below 300 $cm^{-1}$. The Raman modes are assigned according to Naik et al. [30]. The dip at 174 $cm^{-1}$ is the result of interference of three $A_1$(TO) phonon modes, broad band centered at 223 $cm^{-1}$ is the result of $A_1$(TO) phonon mode vibration, a sharp peak at 302 cm$^{-1}$ corresponds to the $B_1$ and $E$(TO + LO) modes, mode at 478 cm$^{-1}$ is assigned as A$_1$(LO) and E(TO), broad asymmetric peak at 520 cm$^{-1}$ is assigned to $E$(TO) and $A_1$(TO) modes. Finally, the peak at 727 cm$^{-1}$ is the result of A$_1$(LO) and E(LO) modes. One can also observe that there are some differences in Raman spectra of BTO and BSTO at lower wave number till 300 cm$^{-1}$. The mode around wave number 223 cm$^{-1}$ is more broaden in BSTO than in BTO. In BSTO, A site is occupied by Ba and Sr atom, this site disorderness relaxes the Raman selection rule and governs the broadening of these modes in BSTO. Also in BSTO, the modes positions below 300 cm$^{-1}$ are shifted towards the lower wave number side. This could be possible due to the reduced average mass by Sr doping at Ba site. This changes the bond strength in BSTO and favors the lower wavelength number in Raman spectra with respect to the pure BTO. Besides these changes there are many more pronounced differences like relative intensity of different modes and a little hump in BSTO around 550 cm$^{-1}$. These differences could be possible due to different temperature regime for different phases in BTO and BSTO, like tetragonal to cubic transition in BSTO is around 340 K, whereas in BTO it is around 403 K. Hence, it is possible that phase fraction of different phases at room temperature in BSTO and BTO would be different.

As we have discussed in earlier section that in BSTO the room temperature phase is not pure tetragonal as in case of BTO but it is the mixture of $P4mm$ and $R\bar{3}c$. We have also explored the Raman spectra of BSTO at different temperatures. Fig. 6(b) illustrates the temperature dependent Raman spectra of BSTO at some selected temperatures from 140 to 370 K. At different temperatures the spectra are shifted vertically for the sake of clarity. From this figure, one can see significant changes in Raman spectra with increasing the temperature. The intensity of most of the modes, except two broad peaks at 260 and 515 cm$^{-1}$ vanishes in the cubic phase i.e. above 335 K. The presence of 260 and 515 cm$^{-1}$ at higher temperatures is ascribed as the higher order Raman scattering processes by Parsons and Rimai [31]. Later on Quittet et al. interpreted these broad bands as the presence of disorder in the paraelectric cubic phase [32]. The evolution of Raman shift of 224 cm$^{-1}$ and 302 cm$^{-1}$ modes with temperature is extracted from these temperature dependent measurements and shown in Fig. 6(c) and (d). From these plots, one can clearly observe that there are significant changes in Raman shift across all the transitions. The Fig. 6 (e) and (f) shows the temperature dependence of the Raman modes. The Raman shift of 302 cm$^{-1}$ mode shows the same temperature dependence as the electric polarization in P-E loops and pyroelectric measurements.

Seo et al. [33] has related the TiO$_6$ octahedral motion with phonons modes in pure BTO. They have shown that 520 cm$^{-1}$ mode is related to octahedral distortion and hump around 630 cm$^{-1}$ is related to the oscillatory stretching motion between the TiO$_4$ plane and apical oxygen atoms. Hence, the evolution of the peak 520 cm$^{-1}$ is the result of interference of different phonon modes with temperature and would provide the information of octahedral (TiO$_6$) distortion and stretching with temperature. Fig. 7(a) shows the temperature dependent Raman spectra from 460 cm$^{-1}$ to 620 cm$^{-1}$. In the tetragonal phase i.e. above 240 K, one can clearly observe that asymmetry and the full width half maximum (FWHM) of 515 cm$^{-1}$ mode has been increased and a weak shoulder like feature appears around 540 to 560 cm$^{-1}$. Fig. 7(b) and (c) shows the FWHM and Raman shift of 515 cm$^{-1}$ mode with temperature, respectively. From Fig. 7(b) and (c), one can clearly see drastic change in FWHM and Raman shift around different transitions which is also consistent with polarization measurement. Pezzotti et al. [34] have reported that a strong asymmetry of the A$_1$(TO$_3$) mode at 515 cm$^{-1}$ is the characteristic of the different orientations of domains in tetragonal BTO.

The different orientation of domains can be possible by TiO$_6$ octahedral distortion and stretching. Marcos et al. [35] have already shown that such type of asymmetry can be induced by the polarized light and described it as domain wall motion induced Raman mode. A careful observation in the dielectric results also shows that there is large frequency dispersion in the real part and tanδ in this temperature regime. Our results corroborate that such feature in Raman spectra coinciding with relaxation behavior in tanδ in BSTO and could be related to the domain walls motion and phase coexistence. From Fig. 7(b) and (c), it is clear that octahedral distortion in tetragonal phase i.e. above 250 K is increasing till 280 K and it leads to the formation of superlattice peak at 280 K as discussed in the SXRD section. This Raman mode does not disappeared completely even at 340 K i.e. tetragonal to cubic transition and leads to P till 350 K in BSTO. Thus our temperature dependent Raman study not only probe all the transitions in BSTO but also infers the changes that create the local maxima in temperature dependent polarization and relaxation behavior in temperature dependent dielectric measurement. If we correlate our temperature dependent Raman study with the relaxation in the dielectric measurement, we realized that it could be possibly due to the different orientation of polarization due to the different phase fraction formed through the $TiO_6$ octahedral distortion. All our temperature dependent studies illustrate the coupling between different order parameters.

## IV. CONCLUSION

To summarize, we have investigated the effect of Sr doping (20 %) in $BaTiO_3$ on its temperature dependent structural, lattice vibrations, dielectric, ferroelectric and electrocaloric properties. The significant enhancement in dielectric constant and remnant polarization is observed for this composition. Our structural analysis corroborate that the room temperature crystal structure of BSTO is not pure *P4mm* but there is additional superlattice peak which is related to $R\bar{3}c$ phase. Moreover, this superlattice peak emerges only in a particular temperature regime i.e. appear around 280 K and disappeared completely at 350 K. The different transitions in BSTO have been illustrated in temperature dependent dielectric, polarization, SXRD, Raman and electrocaloric measurements. The enhancement of dielectric constant, relaxation behavior concomitant with long range ordering and electric polarization are related to the appearance of superlattice peaks around this region. This is most probably associated with the different orientation of ferroelectric domains due to the octahedral distortion in $TiO_6$. Our comprehensive studies on different properties are consistence with each other and establish the correlations in different ordered parameters. The enhancement in ferroelectric properties suggests the potential of such materials for lead free ferroelectrics for different applications.

**ACKNOWLEDGEMENTS**

We would like to thank Arup Kumar Mondel for room temperature XRD measurement, Mr. M. N. Singh for temperature dependent SXRD measurements and Mr. Ajay Kumar Rathore for the help in temperature dependent Raman measurements.

**Figure captions:**

**Fig. 1.** Rietveld refinement of room temperature XRD (a) using *P4mm* space group and (b) with *P4mm* and *R$\bar{3}$c* space groups. Insets in (a) and (b) show the zoomed view near superlattice peak. The arrow in the inset of (a) shows an additional peak which does not belongs to *P4mm* space group.

**Fig. 2.** The temperature dependent (a) real part of dielectric constant (ε′) and (b) tanδ at different frequencies during warming. The inset in (a) illustrates the thermal hysteresis during cooling and warming round tetragonal to orthorhombic transition and inset in (b) shows Arrhenius fitting of the frequency dependent peak. The shaded area in (b) shows the temperature regime where frequency dependent peak is observed.

**Fig. 3.** (a) Isothermal P-E loops of BSTO at some selective temperatures; inset shows the P-E loop of BTO and BSTO at room temperature. (b) Evolution of remnant polarization with temperature and inset shows P-E loops with different maximum applied fields at room temperature.

**Fig. 4.** The temperature dependent (a) remnant polarization, (b) specific heat, (c) ΔT and (d) ΔS of BSTO.

**Fig. 5.** (a) Temperature dependent synchrotron x-ray diffraction of BSTO at some selected temperatures (λ= 0.7525 Å). Inset in (a) shows zoomed view of superlattice peak indicated by arrow. (b) The evaluation of (200) peaks with temperature, arrow indicates the increasing order of temperature. (c) The evaluation of super lattice peak at some selected temperatures. The * represents the superlattice peak in addition to tetragonal phase. The inset in 280 K pattern shows the presence of superlattice peak at this temperature.

**Fig. 6.** (a) Room temperature Raman spectra of BTO and BSTO. (b) Raman spectra of BSTO at different temperatures. (c) and (d) Temperature dependent Raman shifts of broad band centered around 223 cm$^{-1}$ and 302 cm$^{-1}$, respectively. (e), (f) The zoomed view of these modes at different temperatures.

**Fig. 7.** (a) Temperature dependent Raman spectra from 460 cm$^{-1}$ to 620 cm$^{-1}$. (b) The variation of full with half maxima and (c) Raman shift of 515 cm$^{-1}$ mode with temperature.

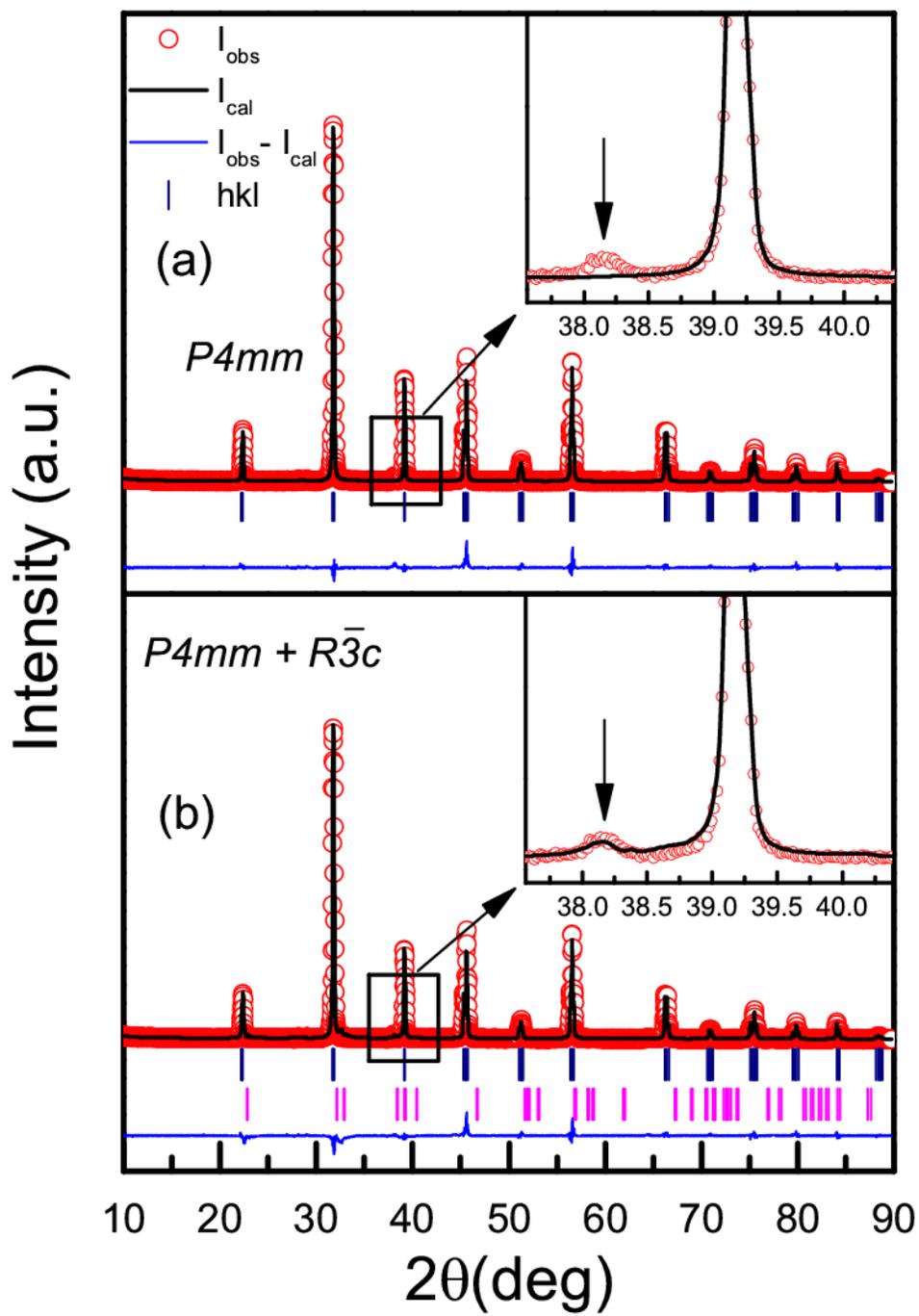

**Fig.1**

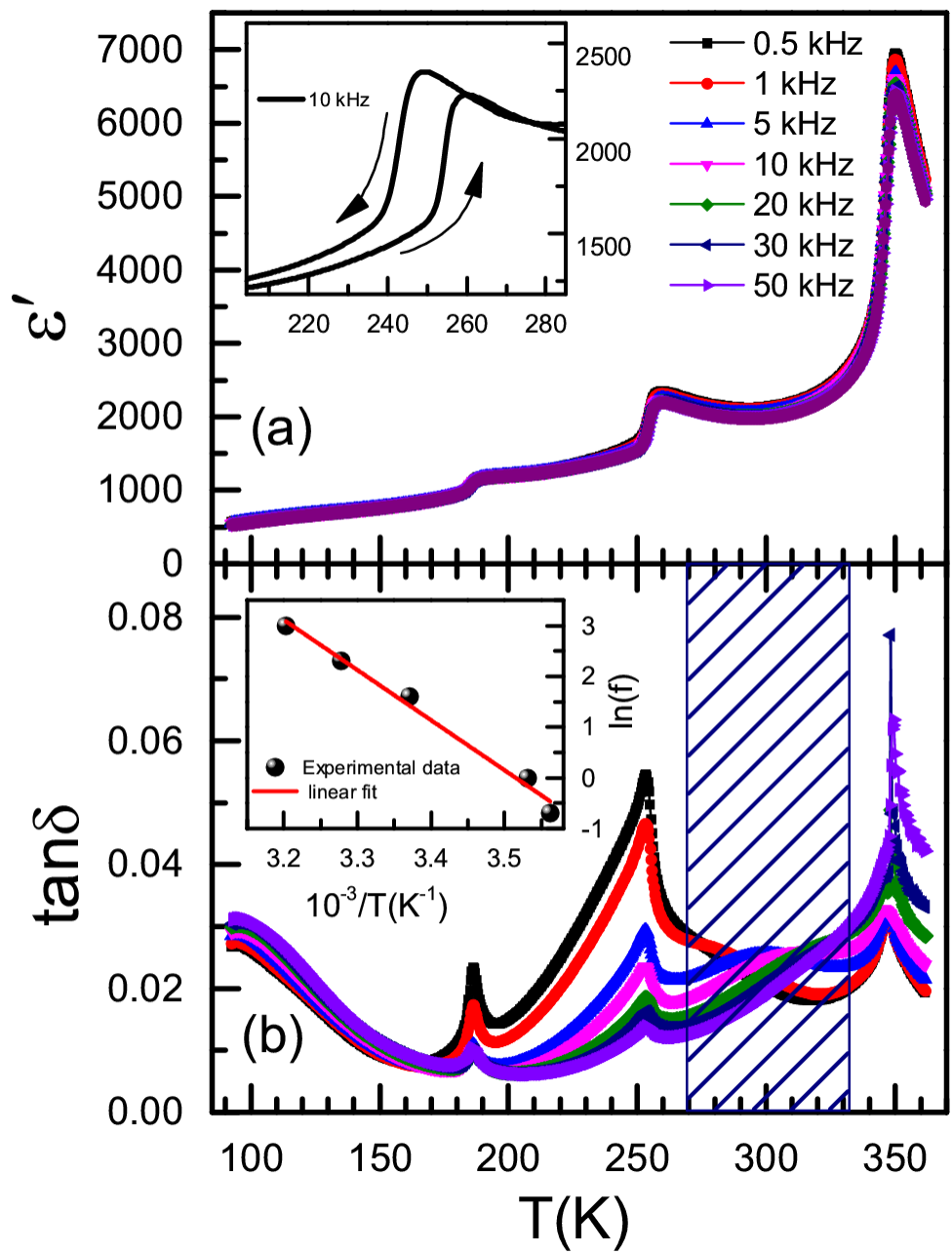

**Fig. 2**

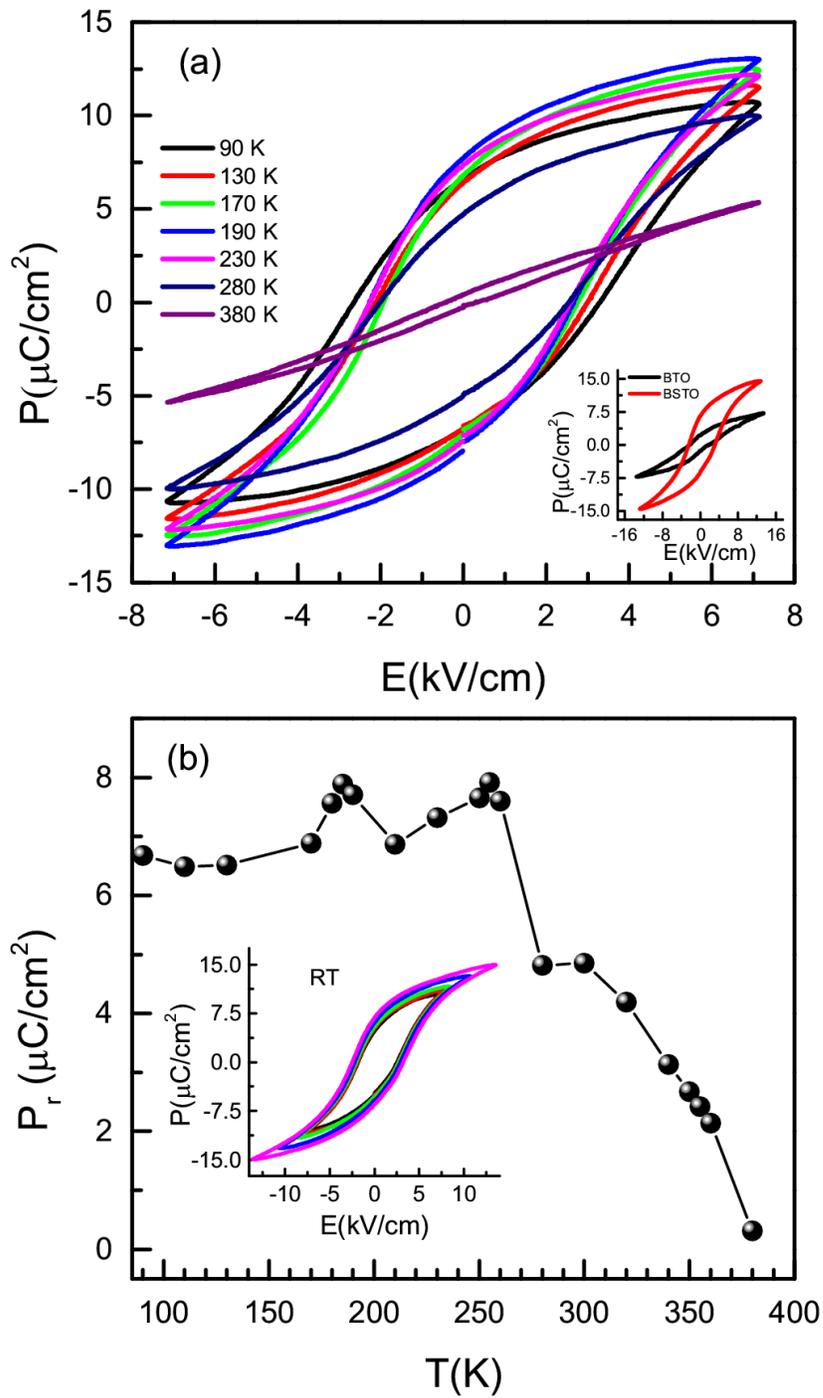

**Fig. 3**

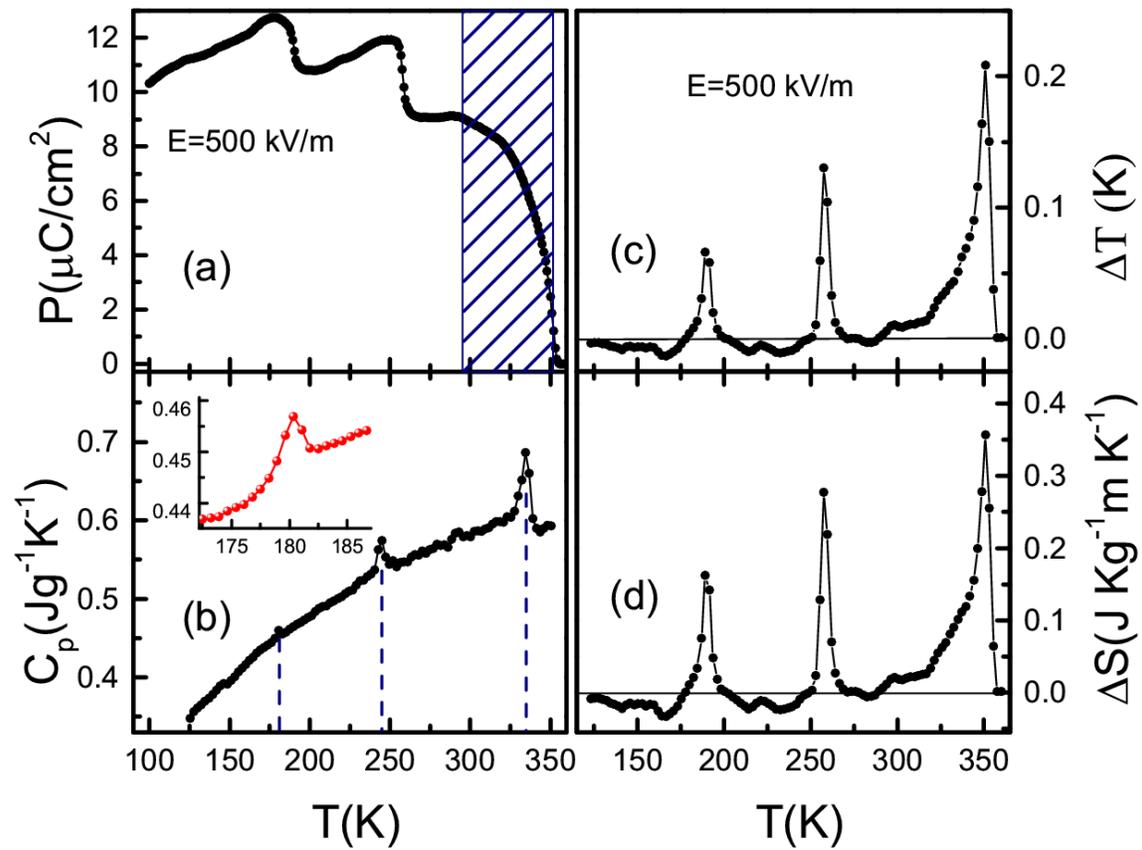

**Fig. 4**

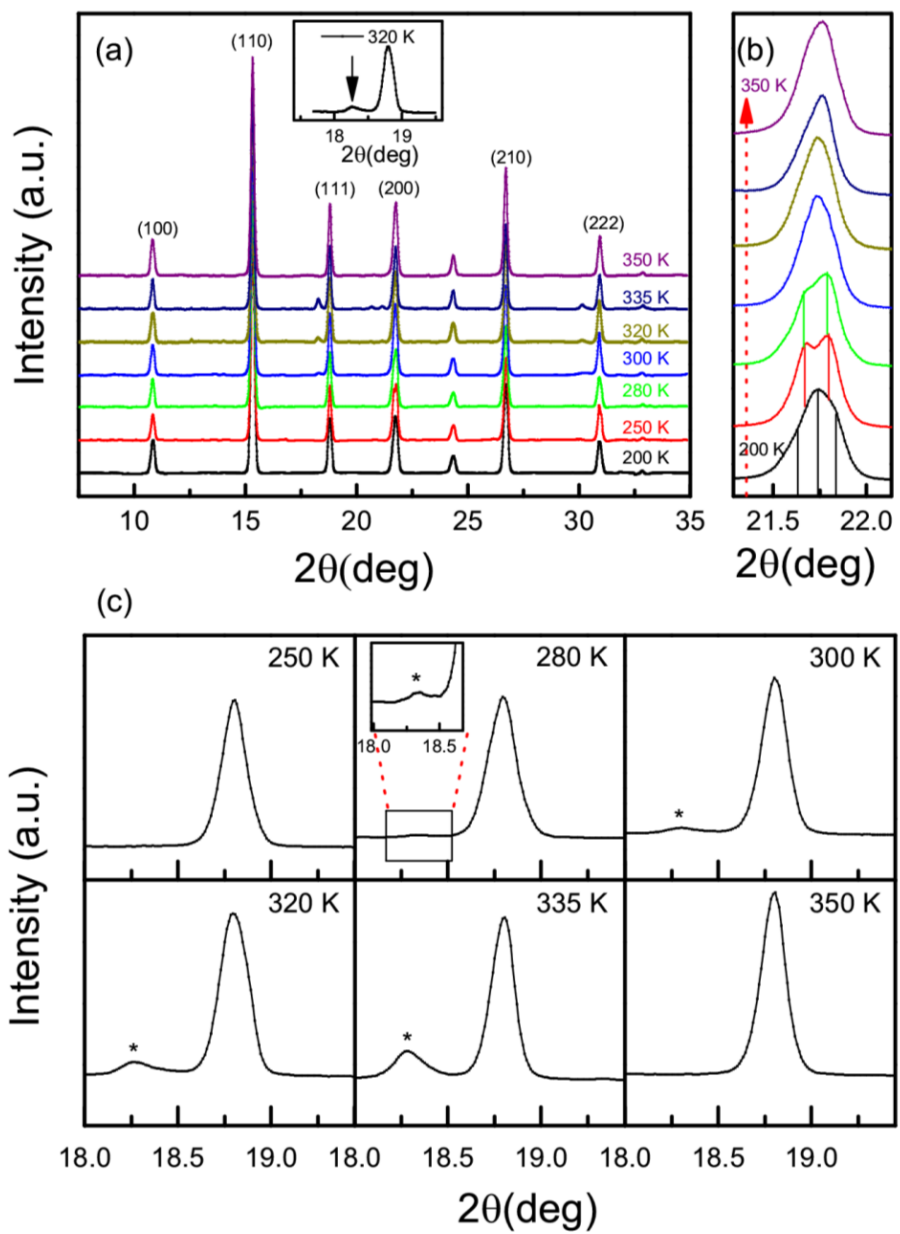

**Fig. 5**

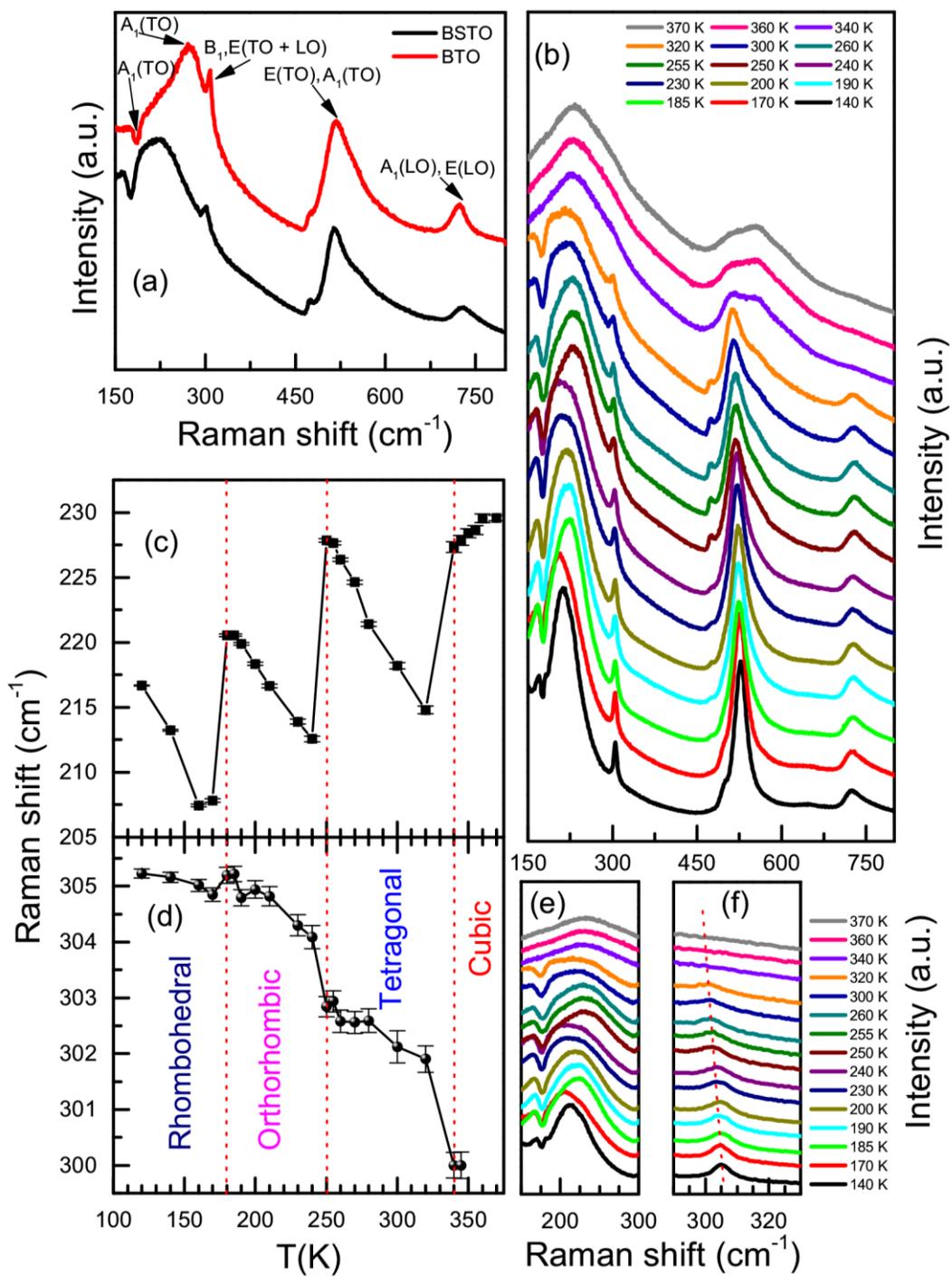

**Fig. 6**

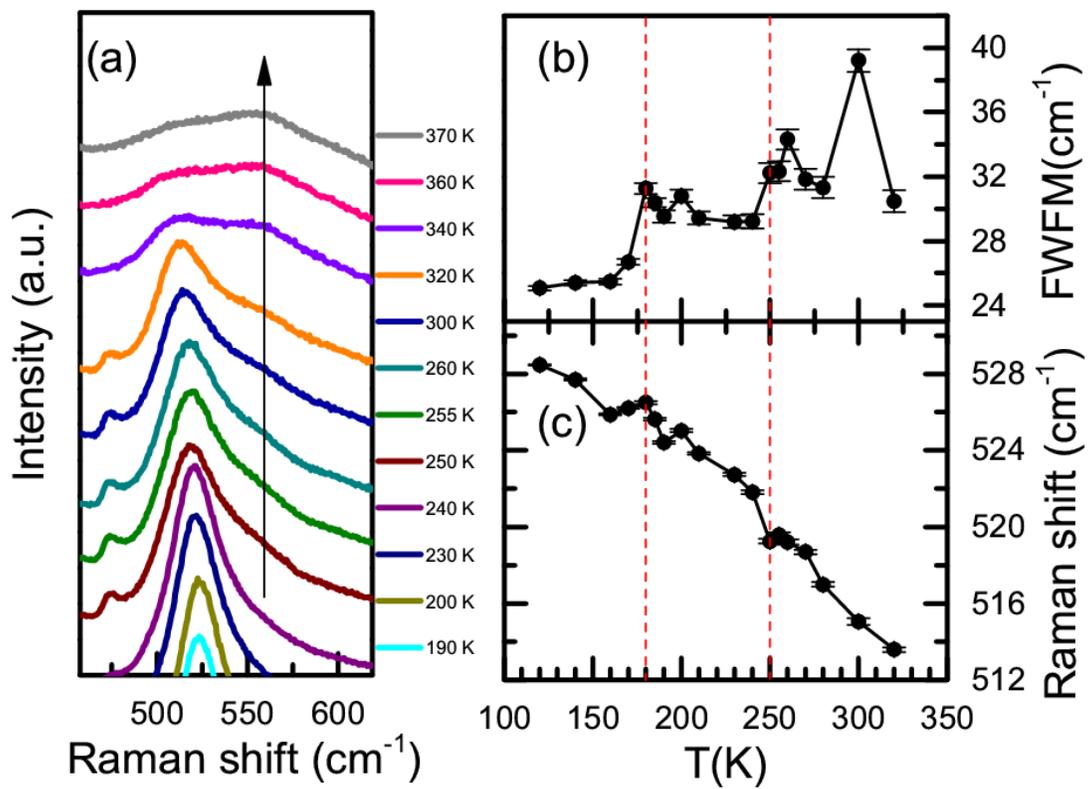

**Fig. 7**